\documentclass[twocolumn]{aastex63}
\usepackage{graphicx}
\usepackage{float}

\submitjournal{PASP}
\shorttitle{Star-Image Centering with Deep Learning II}
\shortauthors{Casetti-Dinescu et al.}

\begin{document}

\title{Star-Image Centering with Deep Learning II: HST/WFPC2 Full Field of View}

\correspondingauthor{Dana I. Casetti-Dinescu}
\email{casettid1@southernct.edu,dana.casetti@gmail.com}

\author[0000-0001-9737-4954]{Dana I. Casetti-Dinescu}
\affiliation{Department of Physics, Southern Connecticut
  State University, 501 Crescent Street, 
New Haven, CT 06515}
\affiliation{Astronomical Institute of the
  Romanian Academy, Cutitul de Argint 5, Sector 4, 
Bucharest, Romania}
\author[0000-0001-5214-7408]{Roberto Baena-Gall\'{e}}
\affiliation{Universidad Internacional de la Rioja,
  Avenida de la Paz, 137,26006, Logro\~{n}o, La Rioja, Spain}
\author[0009-0001-3739-7051]{Terrence M. Girard}
\affiliation{Department of Physics, Southern Connecticut
  State University, 501 Crescent Street, New Haven, CT 06515}
\author[0000-0001-5442-953X]{Alejandro Cervantes-Rovira}
\affiliation{Universidad Internacional de la Rioja,
  Avenida de la Paz, 137,26006, Logro\~{n}o, La Rioja, Spain}
\author{Sebastian Todeasa}
\affiliation{Physics, Applied Physics and Astronomy Department,
  Rensselaer Polytechnic Institute, 110 8th Street, Troy, NY 12180}
\affiliation{Department of Physics, Southern Connecticut
  State University, 501 Crescent Street, 
  New Haven, CT 06515}

\begin{abstract}

  We present an expanded and improved deep-learning (DL)
  methodology for determining centers of 
  star images on HST/WFPC2 exposures. Previously, we demonstrated
  that our DL model can eliminate the pixel-phase bias otherwise
  present in these undersampled images; however
  that analysis was limited to the central portion of each detector.

  In the current work
  we introduce the inclusion of global positions to
  account for the PSF variation
  across the entire chip and instrumental magnitudes to account for
  nonlinear effects such as charge transfer efficiency.
  The DL model is trained using a unique series of WFPC2 observations of
  globular cluster 47 Tuc, data sets comprising over
  600 dithered exposures taken in each of two filters --- F555W and F814W.
  
  It is found that the PSF variations across each chip correspond to
  corrections of the order of $\sim 100$ mpix, while magnitude effects
  are at a level of $\sim 10$ mpix.
  Importantly, pixel-phase bias is eliminated with the DL model;
  whereas, with a classic centering algorithm, the amplitude of this
  bias can be up to $\sim 40$ mpix.
  Our improved DL model yields star-image centers with uncertainties
  of 8-10 mpix across the full field of view of WFPC2.

\end{abstract}


\section{Introduction \label{sec:intro}}

The astrometric potential, for proper-motion purposes,
is yet to be fully
realized for archival images taken with the
Wide-Field Planetary Camera 2
(WFPC2), a legacy instrument of the
Hubble Space Telescope (HST).
Recently we have developed a deep-learning (DL) methodology to
improve the centering precision of images taken with WFPC2.  
These images
are severely undersampled and thus suffer from a fractional pixel bias
in the stars' centers due to the mismatch between the true PSF and
the PSF used by the centering algorithm. To address this problem,
\citet{and2000} built an {\it effective} PSF (ePSF) 
empirically from a set of observations. However, the WFPC2 ePSF library
is not sufficient to remove this pixel-phase bias which can
be as large as 40 to 50 mpix. Once post corrections
such as the 34th-row correction \citep{and1999} and
classic distortion \citep{and2003} are applied,
the pixel-phase bias is manifested as unaccounted-for
noise in the positions. As a consequence, WFPC2
was deemed unfit for high-precision astrometry
and a large archive with images taken between 1993 and 2009
remains untapped for proper-motion studies.

Taking an entirely new approach to determining
stellar centers in WFPC2 images, we developed and refined a
DL code using both simulated and real data
as described in \citet{baena2023} and
in \citet[hereafter, Paper I]{casetti2023}.
At that time, we focused only on the central part of
each WFPC2 chip to avoid the complexity of the PSF
variation across the detector's field of view.
We modeled filters F555W and F814W thanks to a unique data set
available only in these two filters. The method proved
successful, yielding centering uncertainties of the order of
10 mpix per single measurement near the center of each chip.

In the current work, we have expanded upon our previous model
and restructure the DL code to include the variation of the PSF
across the chip as well as magnitude effects in the determination
of new stellar centers. In what follows, we will make frequent
reference to Paper I and encourage the interested reader
to look there for additional, specific details.

\section{Deep Learning Model \label{sec:model}}
\subsection{Model Input \label{subsec:input}}

The data sets we use to build the DL model were taken
in July 1999 at the core of globular cluster
47 Tuc (PID 8267, Gilliland).
These data are unique as there are over 600 exposures
in each filter, taken with fractional-pixel offsets in
each axis,
ranging up to about 2 PC pixels (0.046 ``/pixel).
Theses offsets are 
critical in characterizing the pixel-phase bias.
Conveniently, on the scale of these offsets, effects due
to differential optical distortion
and the 34th-row error can effectively be ignored.
The pattern of offsets is shown in Figure 1 of Paper I;
there is good coverage in both $x$ and $y$ chip coordinates.
There are 636 exposures in F555W and 654 in F814W,
all having 160 seconds per exposure. The data set
is time-wise contiguous,
taken over some 8 days of observations.

We begin by first centering all exposures with the classic ePSF-algorithm
hst1pass\footnote{While using a 2019 version of the code,
we have checked that the 2022 version
\citep{and2022} gives the same results for WFPC2 images;
the ePSF library is identical in the two versions of the code.} code
\citep{and2000,and2022}. 
It is worth noting that this code employs a library consisting of a 
spatial grid of PSFs, meant to properly model its variability across each chip. 
As in Paper I, the hst1pass-determined
positions are used to construct an average catalog in each filter.
This is done by transforming the positions of each exposure into those
of one chosen as reference (typically the first one in the set).
In this way we obtain positions on the same system for all exposures,
which are afterwards averaged with outlier clipping.
The polynomial transformations between target and reference exposures
include up to third-order terms. From this average catalog we eliminate
all objects that have a neighbor within 5 pixels, in order that crowding
effects do not affect the training of the DL model. The average catalog
positions are then used as ``true'' positions in the training process.
The assumption is that these catalog positions are no longer affected by
pixel-phase bias given the large number of nearly-random offsets being averaged
over.

Specifically, the input for the training process
consists of the intensity values of a raster of 
$6 \times 6 $ pixels centered on each star image, along with 
the star's catalog-determined $(x,y)$ center with respect to the 
bottom left corner of the raster.
In addition, we also input the global $X,Y$ coordinates of the
raster within the chip, and the star's instrumental magnitude.
The global-coordinate input will allow the model to compensate
for the variation of the PSF across the field of view,
while the input magnitudes provide allowance for what we suspect
may also be a slight magnitude dependence present in the PSF.
Just to be clear, no attempt is made to have the model predict stellar
magnitudes; we are entirely focused on determining image centers.

The number of input objects per chip and filter are listed in
Table \ref{tab:real-lists}. Compared to Paper 1, we have a factor of
$\sim 9$ more objects, as in that paper we modeled only the central
$1/3$ of the chip. 

\begin{deluxetable}{lccccc}
  \tablecaption{Number of input objects per chip and filter
    \label{tab:real-lists}}
\tablewidth{0pt}
\tablehead{
    \colhead{Filter} &
    \colhead{N$_{exp}$} &
    \colhead{N$_{PC}$} &
    \colhead{N$_{WF2}$} &
    \colhead{N$_{WF3}$} & 
    \colhead{N$_{WF4}$}  
}
\startdata
F555W  & 636 & 2073996  & 1848851 & 2118360 & 1949340   \\
F814W  & 654 & 2378226  & 2290724 & 2524498 & 2371043   \\ 
\enddata
\end{deluxetable}

\subsection{Model Description \label{subsec:description}}

Our specific Convolutional Neural Network (CNN)
model is developed from that presented in
\cite{baena2023} and Paper I.  There, the 
DL model is based on the VGG architecture
\citep{simonyan14}. Our approach does not make 
any assumption regarding PSF shape, which is fundamentally
different from classic centering methods.
Instead, the $(x,y)$ coordinates of stellar centers
are estimated by measuring correlations
with pixel intensity values within a $6 \times 6$
square raster around each star.
\begin{figure*}
    \centering
    \includegraphics[scale=0.60,angle=0]{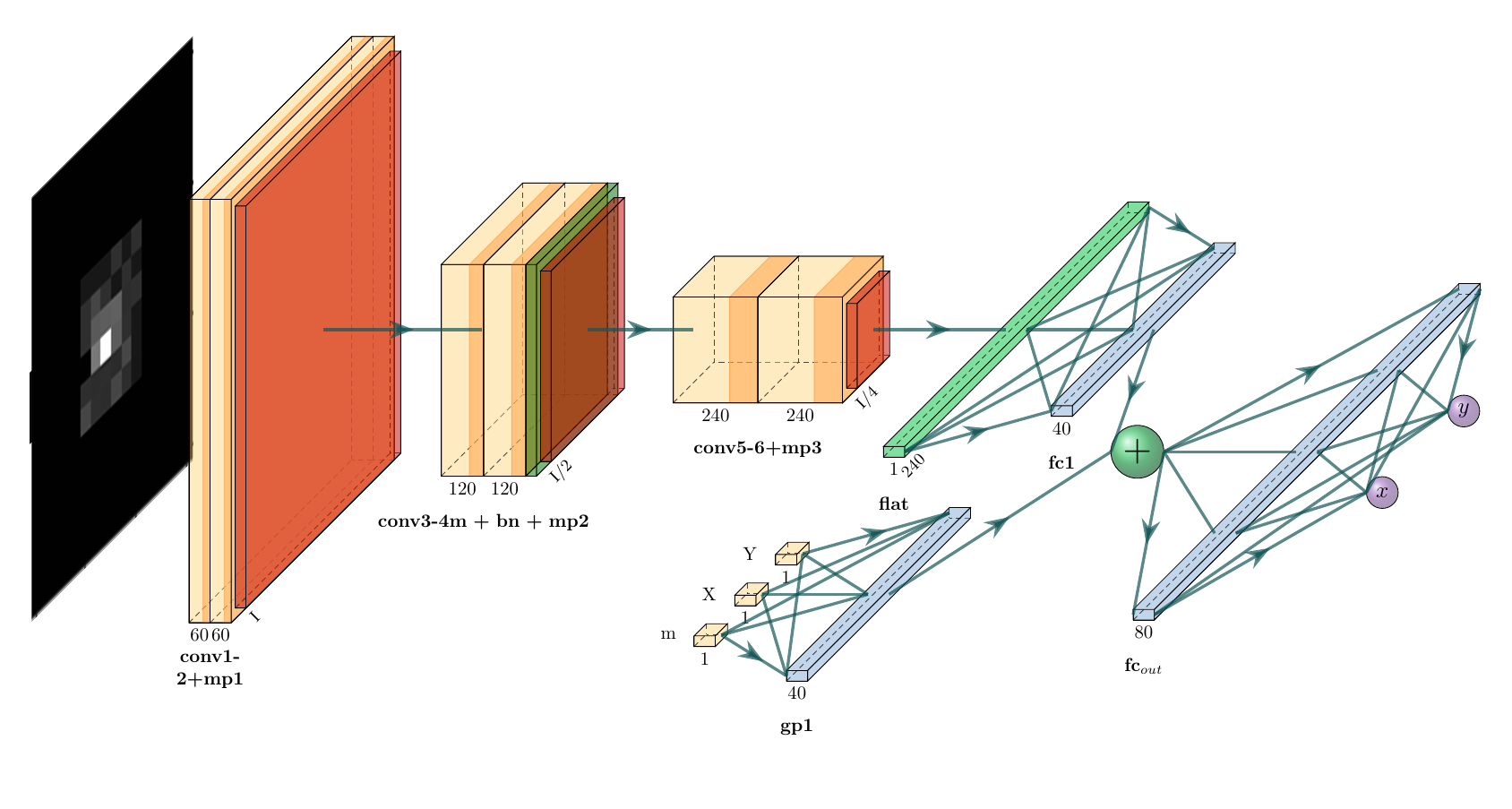}
    \caption{VGG8 model architecture. Input scalars
      are magnitude ($m$) and absolute position ($X$,$Y$).
      Outputs are $x$ and $y$.}
    \label{fig:T10}
\end{figure*}

\begin{deluxetable*}{llrl}
\tablecaption{VGG model \label{tab:vgg-specs}}
\tablewidth{0pt}
\tablehead{
    \colhead{Layer (type)} &
    \colhead{Output shape} &
    \colhead{$\#$ of param.} &
    \colhead{Connected to}
}
\startdata
\hline
conv1$\_$input (InputLayer)  & (None, 12, 12, 1) &  0   & [] \\
conv1 (Conv2D)      & (None, 12, 12, 60) & 1560 & ['conv1$\_$input[0][0]'] \\  
conv2 (Conv2D)      & (None, 12, 12, 60) & 90060 &  ['conv1[0][0]'] \\     
max$\_$pooling2d (MaxPooling2D) &  (None, 6, 6, 60) & 0 & ['conv2[0][0]'] \\             
conv3 (Conv2D)      & (None, 6, 6, 120) & 180120 & ['max$\_$pooling2d[0][0]'] \\ 
conv4 (Conv2D)      & (None, 6, 6, 120) & 360120 & ['conv3[0][0]'] \\
bn$\_$2 (BatchNormalization) & (None, 6, 6, 120)& 480 & ['conv4[0][0]'] \\
max$\_$pooling2d$\_1$ (MaxPooling2D) & (None, 3, 3, 120) & 0 & ['bn$\_$2[0][0]'] \\                
conv5 (Conv2D)    &  (None, 3, 3, 240)   & 259440 & ['max$\_$pooling2d$\_$2[0][0]'] \\                                            conv6 (Conv2D)    & (None, 3, 3, 240)    & 518640 & ['conv5[0][0]' \\                            
max$\_$pooling2d$\_$2 (MaxPooling2D) & (None, 1, 1, 240) & 0 &  ['conv6[0][0]'] \\               
flatten (Flatten)   & (None, 240)  &  0   &  ['max$\_$pooling2d$\_$2[0][0]'] \\                            
gp$\_1\_$input (InputLayer)  &   (None, 3) & 0 &  []     \\
fc$\_$1 (Dense)              &  (None, 40) & 9640 &  ['flatten[0][0]']\\
gp$\_$1 (Dense)              &  (None, 40) & 160  &  ['gp$\_1\_$input[0][0]']   \\
concatenate (Concatenate)   & (None, 80)   &  0   & ['fc$\_$1[0][0]','gp$\_$1[0][0]'] \\
fc$\_$out (Dense)            & (None, 2)   & 162  & ['concatenate[0][0]'] \\
\hline
\multicolumn{4}{l}{Total parameters 1420382; trainable parameters 1420142} 
\enddata
\end{deluxetable*}

The new architecture is illustrated in Figure \ref{fig:T10}.
The model consists of eight trainable layers, six of them convolutional
and the last two fully-connected (FC) layers, with two outputs that
provide paired estimates along the $x$- and $y$-axis, respectively.
Hence, it should be considered a VGG8 model. One maxpool layer is
inserted every two convolutional ones, and all hidden layers
are equipped with Rectified Linear Unit (\emph{ReLu})
non-linear activation functions
except the last one, which is \emph{linear}, in accordance with
the regression nature of this problem. As in Paper I, inserting
a batch-normalization layer after the fourth convolutional one
helps stabilize the convergence process in our specific problem.

As an important novelty with respect to Paper I,
scalar parameters, such as the star magnitude $m$ and its
global position within the CCD ($X$ and $Y$), are fed into an
FC layer and then concatenated to the features extracted along
the main branch before the last fully connected layer.
The underlying idea is that the main branch computes estimates
based on correlations between the intensity values within
the raster, while the scalar parameters perform a fine
tuning informing the network about possible dependencies of
the PSF shape with respect to the star magnitude
and its position across the CCD. 

The design process of the model also differs with respect to
the aforementioned VGG6 model in Paper I in two important aspects.
Firstly, in Paper I the architecture and the
hyperparameter space values
(i.e., number of layers and kernels within, kernels sizes,
number of batches and epochs, type of optimizer, loss function, etc.)
were derived from a set of $\sim 4000$ mock stars in each chip.
The model was then trained over a data set of real 
stars in WFPC2 images. Therefore, although the simulator used
to create the mock images was based on the WFPC2 ePSF library
of \citet{and2000}, one may expect an inconsistency between
the dataset used to design the hyperparameter space of the model
and the dataset used to train, validate and test it.
In the current work, for the sake of consistency of
the model at both design and training stages, we
used real star images throughout.
Specifically, at the design stage
we used a subset of $\sim 12000$ star images
(from the F555W PC chip).

Secondly, we have performed an automatic search across
the hyperparameter space of solutions to find the
best combination. For this purpose,
the \emph{HalvingRandomSearchCV} estimator was used,
which searches within a parameter space using successive
halving \citep{Li2018}. This is somewhat like a competition
between different candidate combinations. It is an
iterative selection process where all candidates ---
i.e., the hyperparameter combinations --- are evaluated using
a small amount of resources in the first iteration.
Those combinations considered the best are
selected for the next iteration,
which are then allocated more resources. Hence, the number of
resources are increased as the number of candidates decreases
until finding the best combination of hyperparameters.
Other estimators make use of grid parameter search strategies,
which are more exhaustive
but more expensive in terms of computational burden. 

After exploring several hundreds of hyperparameter combinations
with this tool, we found our best design with the
following hyperparameters: $6$ convolutional layers,
$60$  kernels in the first group of two, doubling the number
of kernels as the model increases in depth. Both the
flattened output of the convolutional part and the three
scalar inputs are processed by FC layers of $40$ neurons,
concatenated and passed to the output FC layer of $80$ neurons.
Thus, the network has a total of $\sim 1.4$M trainable
parameters. The properties of the model are
sown in Table \ref{tab:vgg-specs} and 
in Figure \ref{fig:T10}. The same architecture
is used for all four WFPC2 chips and both filters. 

For comparison, the VGG6 model in Paper I
was made up of $\sim 0.2$M trainable parameters.
The difference between that model and the new VGG8 
illustrates the increase in complexity. The
new model must account for the variation of the PSF
across the chip, as well as the variation of the PSF
with magnitude.

We also obtained a best learning rate of $4 \times 10^{-5}$,
and a weight decay $8 \times 10^{-3}$. Different loss functions
and metrics for validation were tried, such as the
\emph{logcosh}, \emph{huber\_loss}, or \emph{cosine\_similarity},
and the \emph{MeanAbsoluteError}. The last of these was found
to be the most efficient (as in Paper I).
Finally, different types of
optimizer were also tested such as \emph{Adam}, \emph{Nadam},
or \emph{RMSProp}. We found \emph{AdamW} provided the best
solution with an Exponential Moving Average (EMA)
momentum equal to $0.95$. The input
data set of stars was divided into subgroups of $70 : 10 : 20\% $
for training, validation, and final testing, respectively.
We did not find noticeable differences with respect to the
batch size. All input cutout images’ intensities are first normalized
to a sum of one.
The initial $6 \times 6$ raster images
are zero-padded, thus increasing their size
to $12 \times 12$. This guarantees the relevant information
(i.e., pixel intensities)
within the raster is preserved
as the network layers reduce their sizes due to edge
effects after convolutions with the kernels.
The model was designed
in Keras/TF \citep{chollet2018}.

\subsection{Model Outcome \label{subsec:outcome}}

\begin{figure}[H] 
\includegraphics[scale=0.51,angle=0]{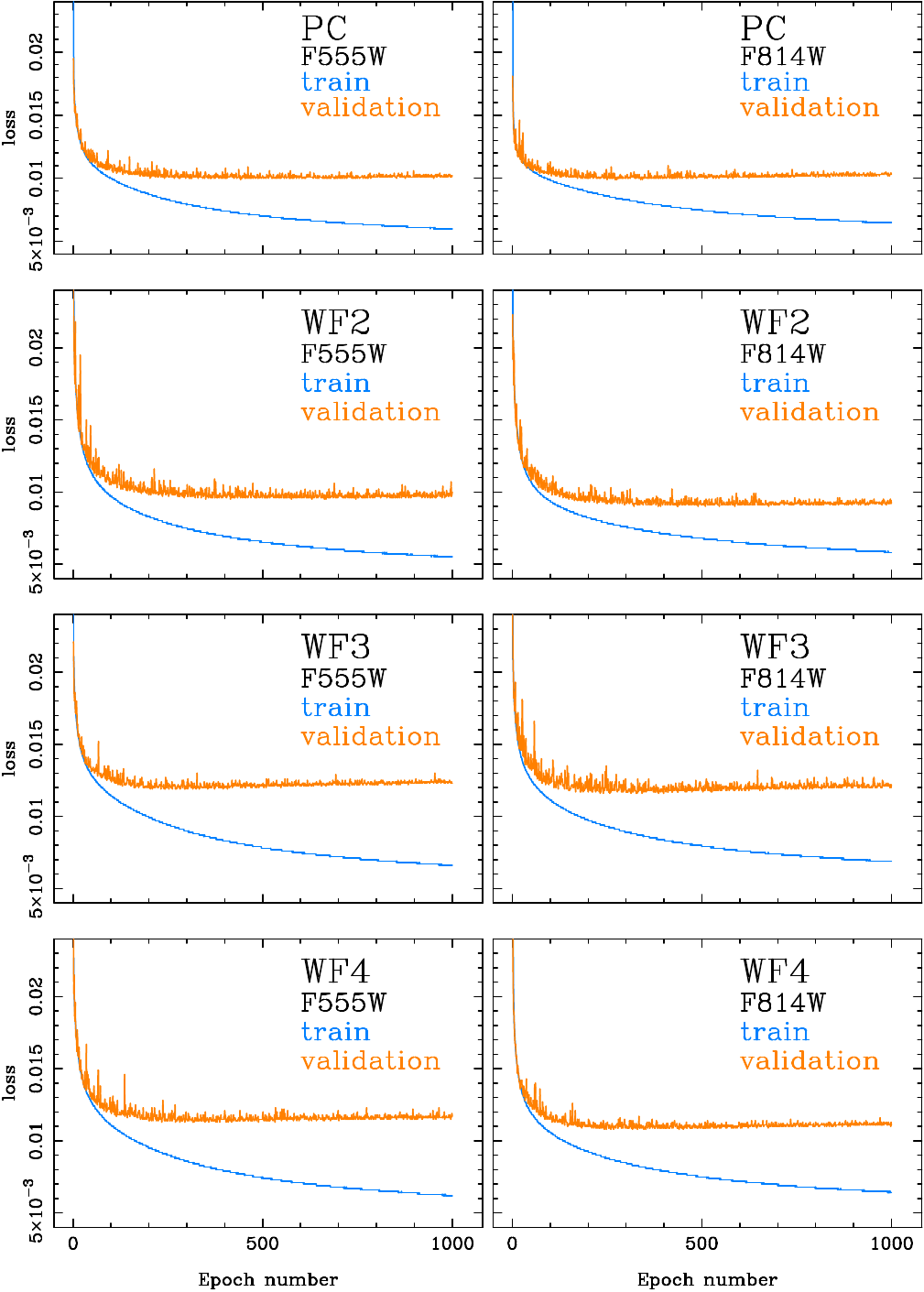}
\caption{The loss function trend with
  epoch for the four chips and the two filters. Training and
validation samples are shown.}
\label{fig:loss}
\end{figure}

In Figure \ref{fig:loss} we show the loss function
for the training ($70\%$) and the
validation ($10\%$) samples  
as a function of epoch during the training process.
We have experimented with a couple of values for the
number of epochs, and eventually settled for
1000. The curves indicate that no overfitting is present.

Next, we look at the test sample which
consists of $20\%$ of the input objects
listed in Tab. \ref{tab:real-lists}. We calculate the
standard deviation of the
differences between output and input positions for
the test sample and list these in Table \ref{tab:stddev-test}.
To calculate these standard deviations, we limit the
test sample to a magnitude range corresponding to
well measured stars, which is the same for all chips and filters.
We also discard $3\sigma$ outliers. The values in Tab. \ref{tab:stddev-test}
represent the centering error per single measurement in
detector millipixels (0.046 ``/pixel for the PC and 0.10 ``/pixel for WF chips).
Note these are conservative estimates, as they assume the input catalog positions
are completely free of error.
\begin{deluxetable}{lcccc|ccc}
  \tablecaption{Standard deviations of position
    differences and number of objects in each test sample
    \label{tab:stddev-test}}
\tablewidth{0pt}
\tablehead{
  \multicolumn{1}{l}{Chip} & \multicolumn{3}{c}{F555W} & & \multicolumn{3}{c}{F814W} \\
  \hline
  & \multicolumn{1}{c}{$\sigma_x$} & \multicolumn{1}{c}{$\sigma_y$} & \multicolumn{1}{c}{$N_{obj}$} & &
  \multicolumn{1}{c}{$\sigma_x$} & \multicolumn{1}{c}{$\sigma_y$} & \multicolumn{1}{c}{$N_{obj}$} \\
  & \multicolumn{2}{c}{(mpix)} & & & \multicolumn{2}{c}{(mpix)} & 
}
\startdata
PC & 8.6 & 9.1 & 367955 && 9.2 & 9.6 & 427849  \\
WF2 & 9.0 & 8.9 & 233466 && 7.8 & 7.7 & 395907  \\
WF3 & 8.7 & 8.7 & 331569 && 9.2 & 9.0 & 404576  \\
WF4 & 8.2 & 8.6 & 312578 && 8.3 & 8.7 & 388009  \\
\enddata

\end{deluxetable}
\section{Applying the DL Model  \label{sec:results}}
\subsection{Cluster Results \label{subsec:clusters}}
The models derived in Sec. \ref{sec:model} are now applied
to the (hst1pass-determined) detections in all exposures of the NGC 104
data set, as well as to other cluster data sets that have not participated
in the building of the model. We require data sets that are
rich in well-measured stars and have repeated exposures with small
offsets, such that other possible systematic errors operating on
scales of tens of pixels do not affect the analysis.

We use the same cluster data sets as in Paper I, as these have already been
deemed the most appropriate for such testing. The cluster
fields, per filter, are listed in Table \ref{tab:data-sets},
including the number of exposures, exposure times,
and the epoch of observation. In the last column of
Tab \ref{tab:data-sets} we add a data set number that
corresponds to the specific MAST DOI link listed in the
Acknowledgments.

Star detection and cutout rasters are made using hst1pass star
centers as preliminary position estimates. Then, the
DL models, by filter and chip, are applied to each of these sets to
calculate new star centers.

To evaluate these, 
polynomial transformations of star positions are made for each
exposure into a chosen reference exposure
and the standard errors of these transformations are recorded.
The standard errors are then plotted as a
function of the pixel phase of the offset from the
reference exposure. We denote this pixel phase of the offset
$\phi$, and its value ranges from 0 to 1. In other words,
this is the fractional part of the full offset.
In such plots, the pixel-phase bias shows itself as a
curve with minimum standard error at $\phi$ = 0 and 1,
but rising to an elevated level at mid values of $\phi$.
Conversely, when no pixel-phase bias is present,
the standard error curves will be flat.
The entire process is repeated using the original hst1pass
centers, in order to compare the amount of pixel-phase bias present
in the two centering algorithms.
\begin{deluxetable}{lrrcc}
  \tablecaption{Data sets used to test the DL models
    \label{tab:data-sets}}
\tablewidth{0pt}
\tablehead{
    \colhead{Target field} &
    \colhead{N$_{exp}$} &
    \colhead{Exp. time} &
    \colhead{Epoch} &
    \colhead{Data Set \#}
}
\startdata
\multicolumn{4}{c}{F555W} \\
\hline
\bf{NGC 104}  & {\bf 636} & {\bf 160}  & {\bf 1999.5} & {\bf 1}   \\
NGC 6752 - PC & 118 &  26  & 1994.6   & 3 \\ 
NGC 6441      &  36 & 160  & 2007.3   & 4 \\ 
NGC 6341      &  28 & 100  & 2008.1   & 5 \\
\hline
\multicolumn{4}{c}{F814W} \\
\hline
\bf{NGC 104}  & {\bf 653} & {\bf 160}  & {\bf 1999.5} & {\bf 2} \\
NGC 6752 - PC & 109 &  50  & 1994.6   & 6 \\
NGC 6656      & 162 & 260  & 1999.1   & 7 \\ 
NGC 6205      &  25 & 140  & 1999.8   & 8 \\ 
NGC 5139      &  24 &  80  & 2008.1   & 9 \\ 
\enddata
\end{deluxetable}

\begin{figure}[H]
\includegraphics[scale=0.47,angle=0]{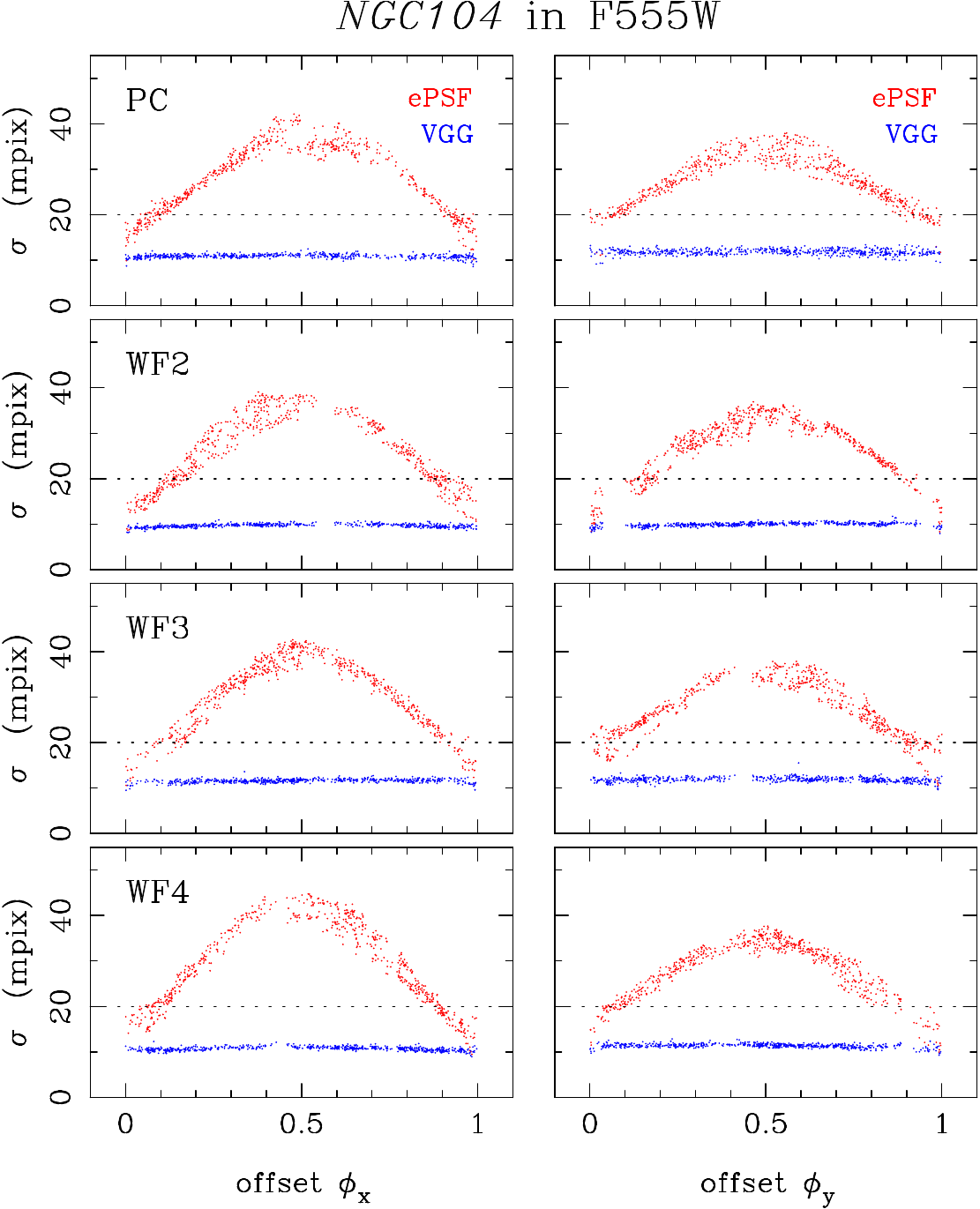}
\caption{Standard error of the transformation
  (of a target exposure into the reference exposure) as a
  function of offset phase. Each row represents a chip while
  $x-$ and $y-$coordinate values are shown in the
  left and right panels, respectively.
    Larger standard errors near mid-pixel phase indicate the 
  presence of pixel-phase bias in the positions.
  The classic ePSF/hst1pass centering algorithm is shown with red symbols,
  while the DL algorithm is shown with blue symbols.
  A benchmark level of 20 mpix is shown with a dotted line. 
  Note the flat curves obtained for the DL centers at $\sim 10$ mpix.
  These results are for exposures of NGC 104 in filter F555W.
}
\label{fig:f5-ngc104}
\end{figure}

The results for NGC 104 in F555W
presented in Figure \ref{fig:f5-ngc104}
show a dramatic improvement of the DL
positions over those obtained with hst1pass,
in all four chips. This demonstrates that
the method works, not only for the central part of the chip
as shown in Paper I, but also across the full field of view of each chip.
Our refinement of the DL model 
has effectively allowed for modeling of
the PSF variation across the chip.

\begin{figure}[H] 
\includegraphics[scale=0.47,angle=0]{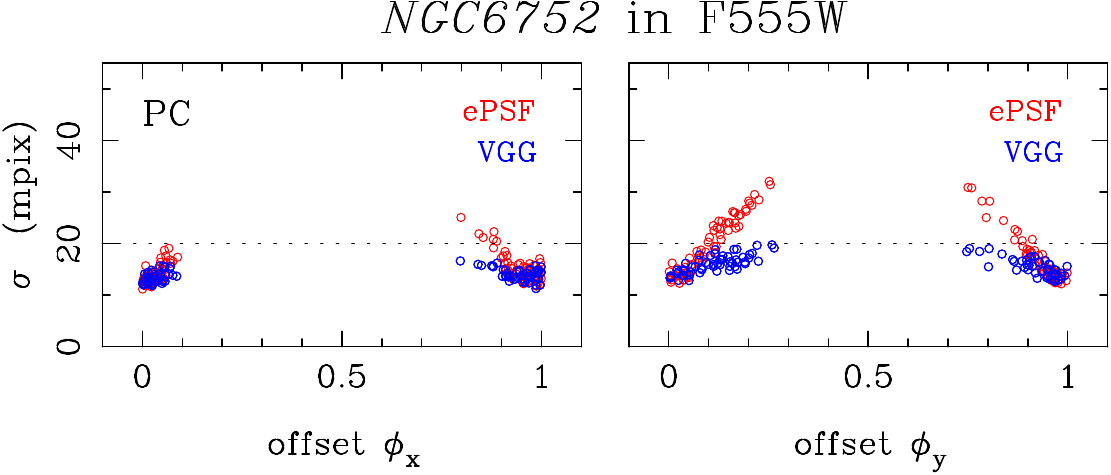}
\caption{As in Fig. \ref{fig:f5-ngc104}, only for NGC 6752 in F555W.
  The phase coverage is poor in this case, however the trends of
  the curves are apparent.}
\label{fig:f5-ngc6752}
\end{figure}

\begin{figure} 
\includegraphics[scale=0.47,angle=0]{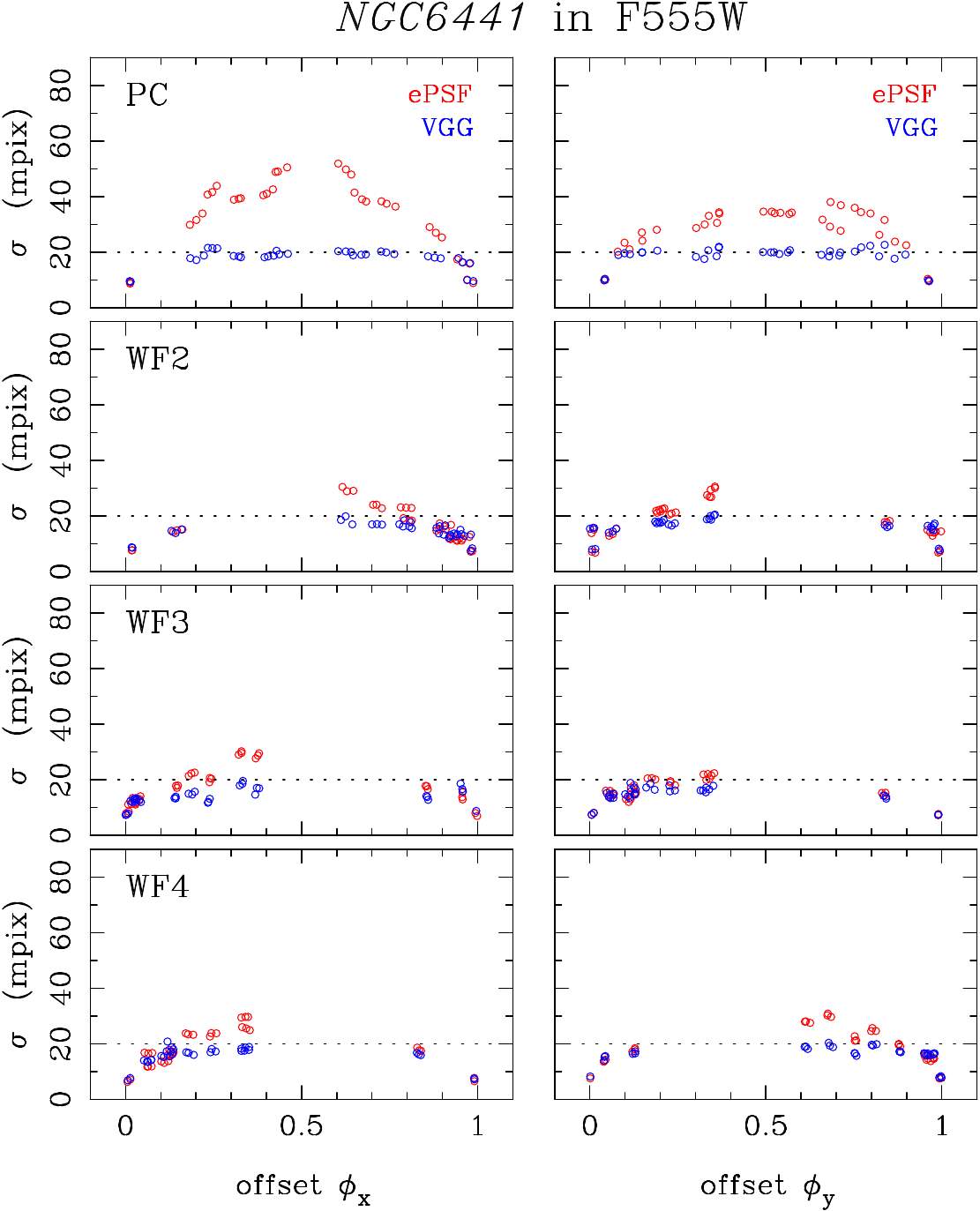}
\caption{As in Fig. \ref{fig:f5-ngc104}, only for NGC 6441 in F555W.}
\label{fig:f5-ngc6441}
\end{figure}

\begin{figure} 
\includegraphics[scale=0.47,angle=0]{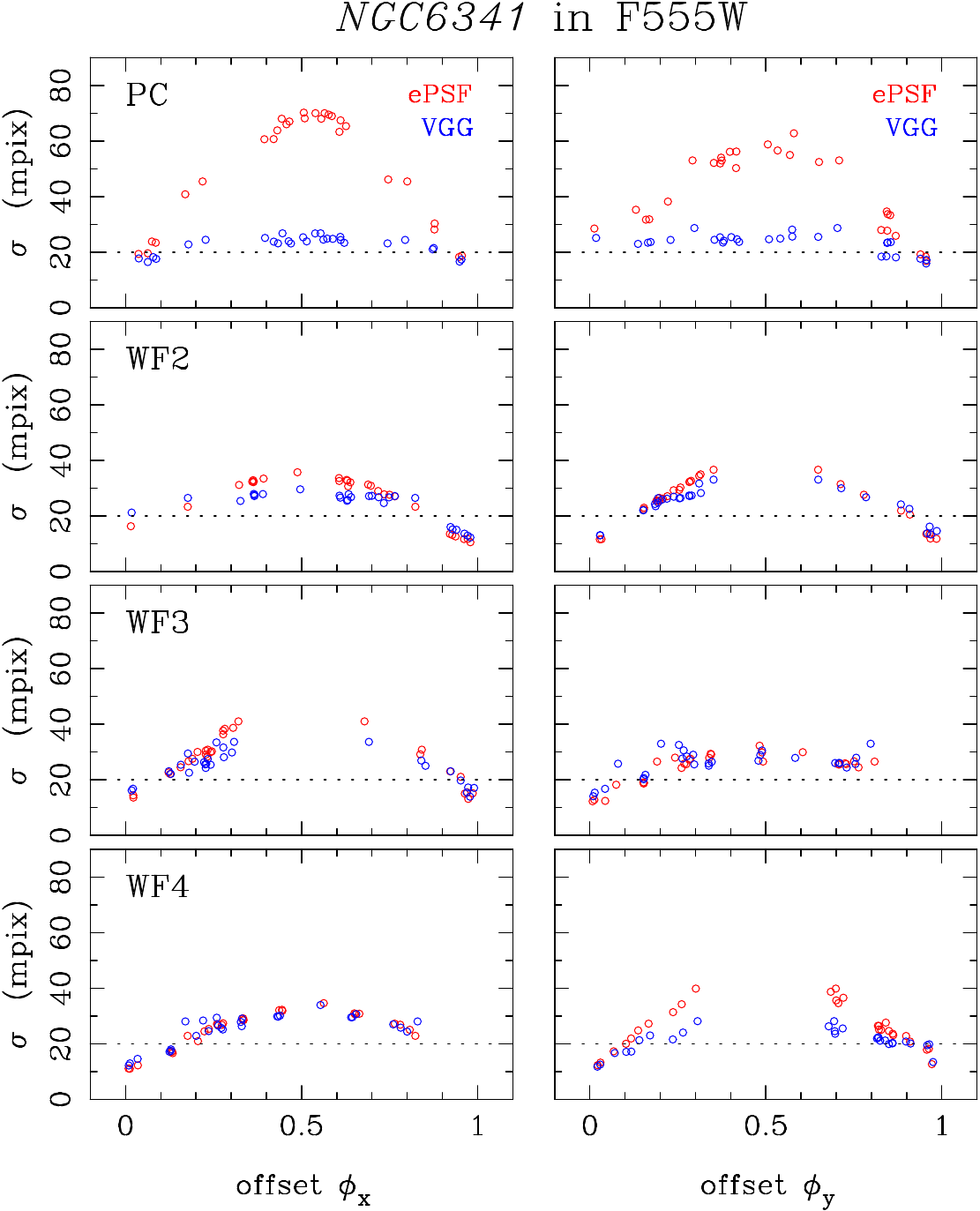}
\caption{As in Fig. \ref{fig:f5-ngc104}, only for NGC 6341 in F555W.
  This data set is the farthest in time with respect to the
  NGC 104 set on which the DL model is trained.
  Therefore, long-term time variation of the PSF may be
responsible for the lesser improvement of image centers in the WF chips.}
\label{fig:f5-ngc6341}
\end{figure}

Results for the other
data sets observed in filter F555W are shown in Figures
\ref{fig:f5-ngc6752}, \ref{fig:f5-ngc6441}  and
\ref{fig:f5-ngc6341}.

Note that the DL models applied to these sets were
trained on ($70\%$ of) the NGC 104 data,
while the target sets were taken at a
different observation epoch, at which temporal changes in
the PSF may become important.
Also, none of these sets match the number of repeats and 
offset-phase $\phi$ coverage of the NGC 104's data sets.
Therefore, the standard error plots appear less coherent.
Lastly, variation in the observation
exposure times and richness of these other target clusters
yield a large range in quality for their
standard-error plots.

The data set for NGC 6752 had only PC
observations, in both filters. In spite of the
poor sampling in offset pixel phase, the
trends are apparent: the DL positions
are less affected by the bias error compared to the
hst1pass positions --- see Figs \ref{fig:f5-ngc6752} and
\ref{fig:f8-ngc6752}.

Results for clusters NGC 6441 and NGC 6341 shown in
Figures \ref{fig:f5-ngc6441} and \ref{fig:f5-ngc6341}
show great improvement
for the PC, but more modest improvement for the WF chips.
Since these observations are taken some 8 to 9 years after the
NGC 104 set, it is possible that the PSF of the WF chips varied
with time, and thus the NGC 104-based model is less
representative for these two clusters.


\begin{figure}[H]
\includegraphics[scale=0.47,angle=0]{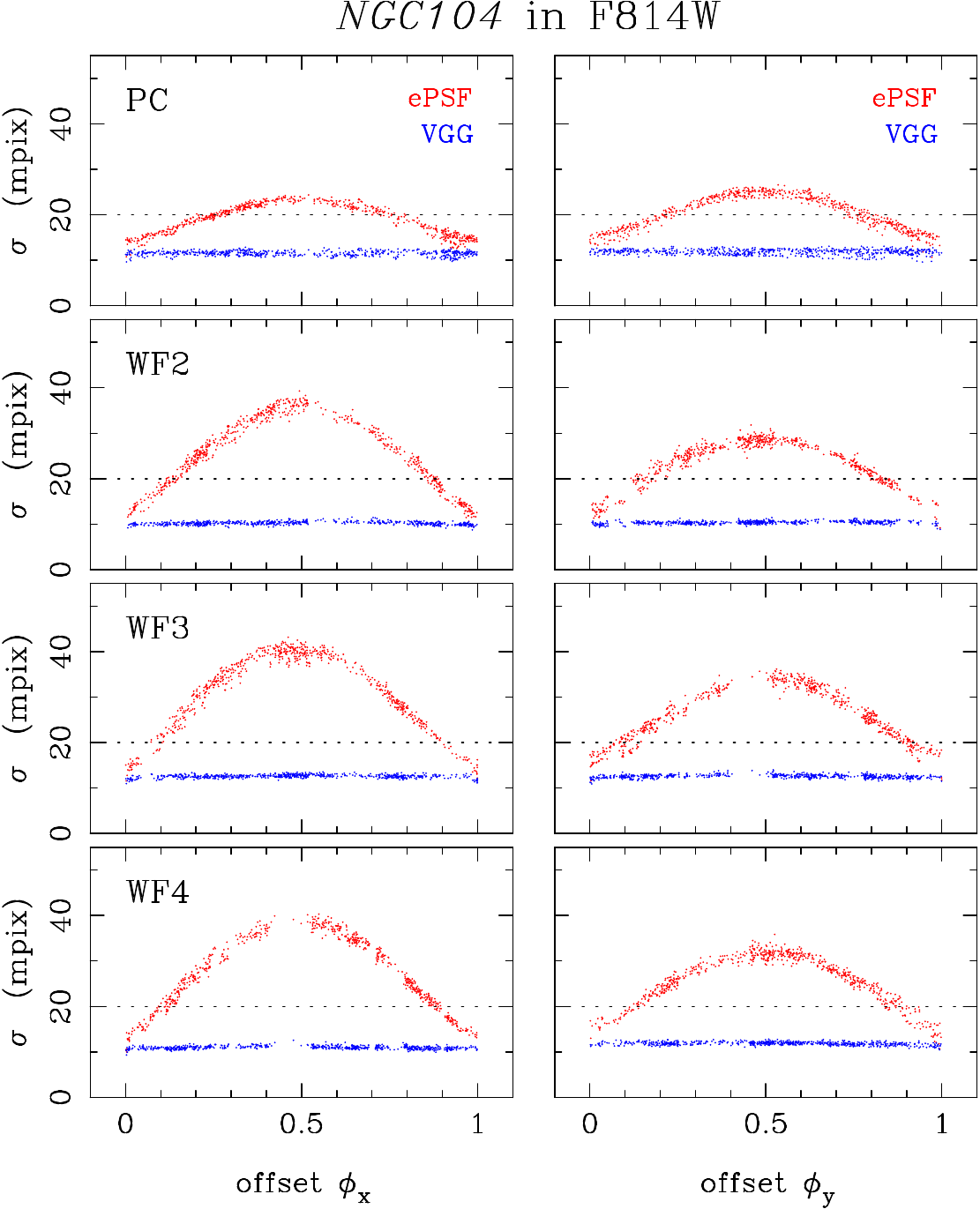}
\caption{As in Fig. \ref{fig:f5-ngc104}, only for NGC 104 in F814W.}
\label{fig:f8-ngc104}
\end{figure}

\begin{figure}[H]
\includegraphics[scale=0.47,angle=0]{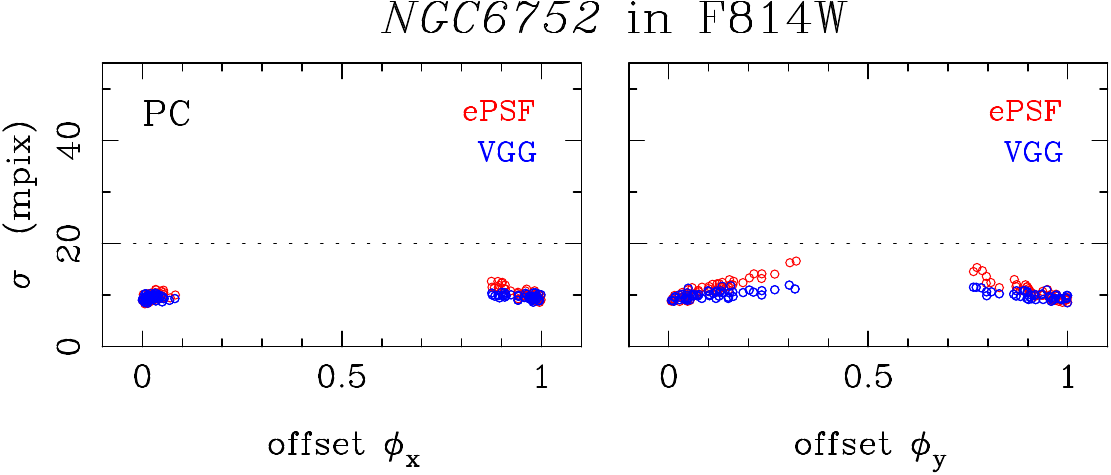}
\caption{As in Fig. \ref{fig:f5-ngc104}, only for NGC 6752 in F814W.}
\label{fig:f8-ngc6752}
\end{figure}

\begin{figure}[H]
\includegraphics[scale=0.47,angle=0]{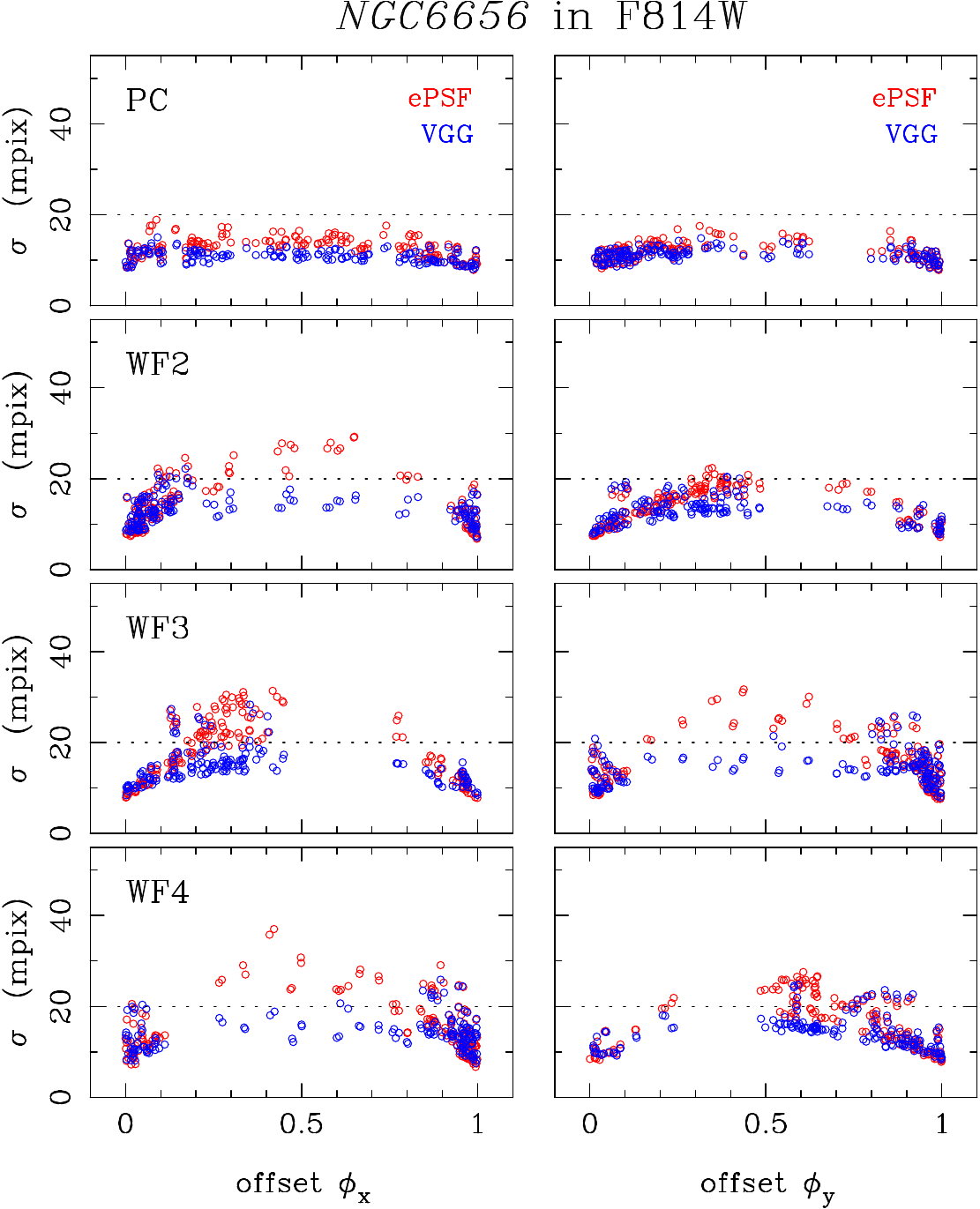}
\caption{As in Fig. \ref{fig:f5-ngc104}, only for NGC 6656 in F814W.}
\label{fig:f8-ngc6656}
\end{figure}

\begin{figure}[H]
\includegraphics[scale=0.47,angle=0]{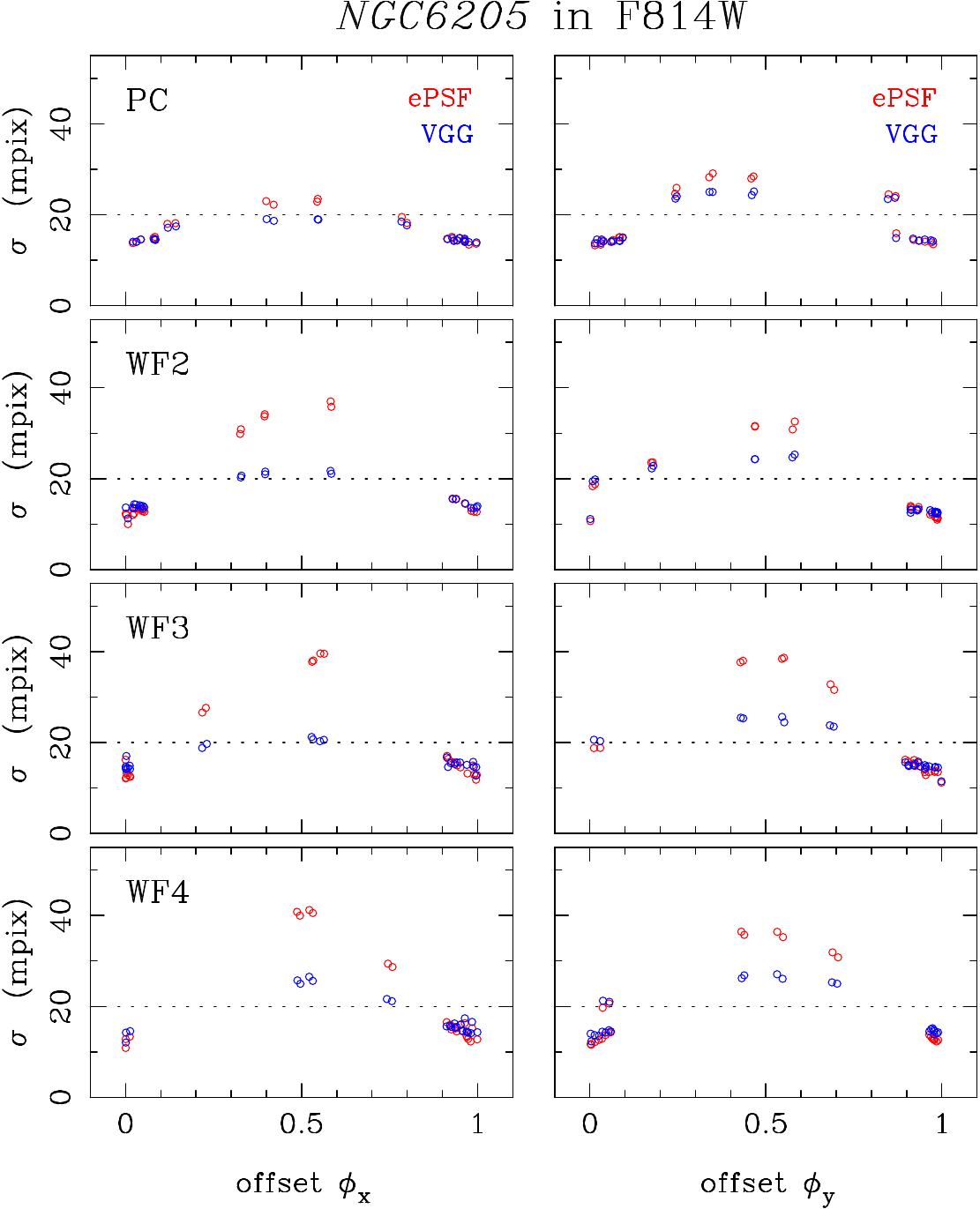}
\caption{As in Fig. \ref{fig:f5-ngc104}, only for NGC 6205 in F814W.}
\label{fig:f8-ngc6205}
\end{figure}

\begin{figure}
\includegraphics[scale=0.47,angle=0]{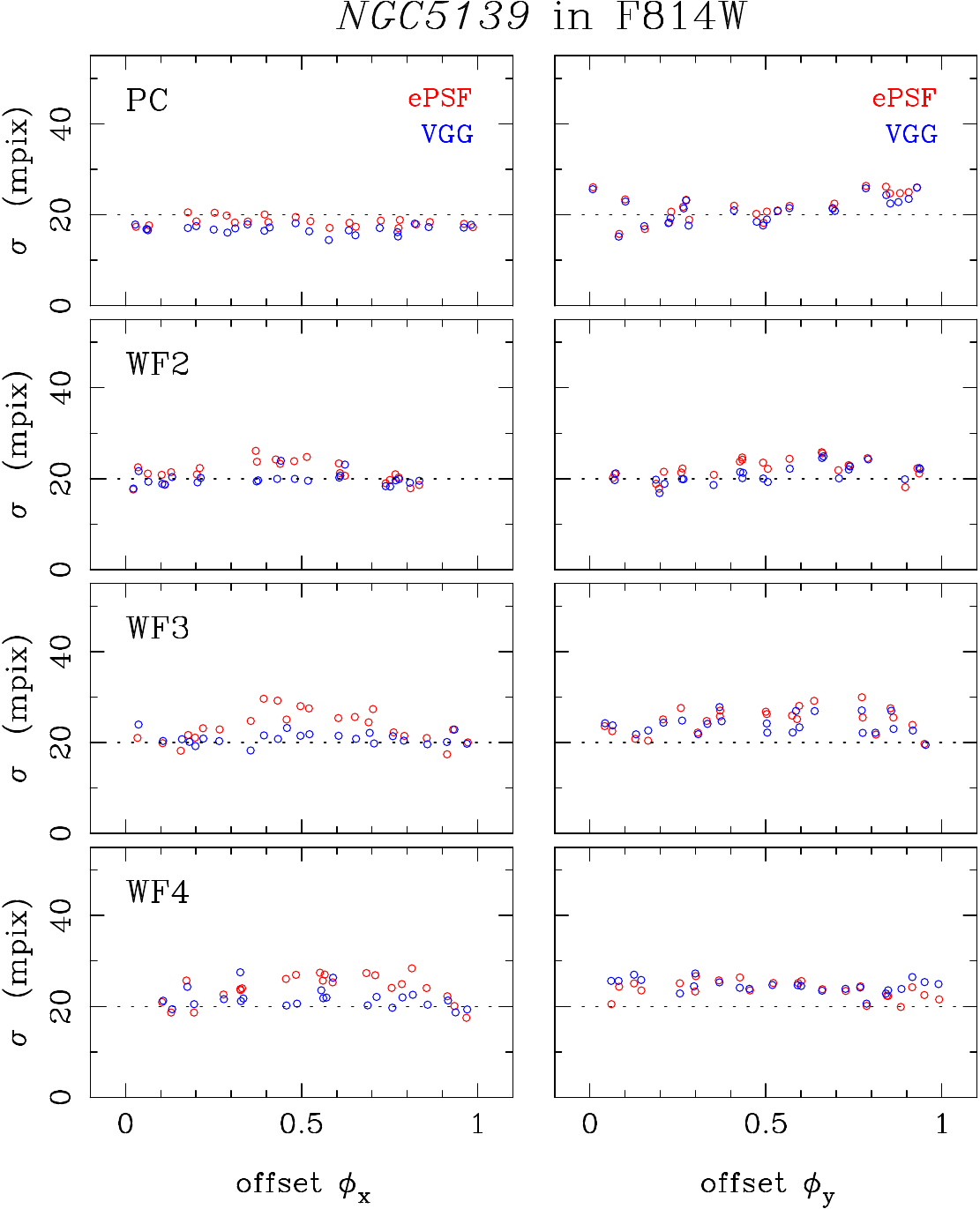}
\caption{As in Fig. \ref{fig:f5-ngc104}, only for NGC 5139 in F814W.}
\label{fig:f8-ngc5139}
\end{figure}

In filter F814W, the NGC 104 results once again show
great improvement in removing the pixel-phase bias
(see Figure \ref{fig:f8-ngc104}). Results for
NGC 6656 and NGC 6205 shown in Figures \ref{fig:f8-ngc6656}
and \ref{fig:f8-ngc6205} are good for the WF chips and more modest for the PC.
Finally, for cluster NGC 5139 there is hardly
any improvement of the DL positions over the hst1pass ones as
seen in Figure \ref{fig:f8-ngc5139}.
Overall, this data set has large errors, in excess of 20 mpix,
and the curves are rather flat, hardly indicating
a pixel-phase bias. It is possible that the
signal-to-noise ratio dominates the errors in this case.
Therefore, we regard this data set as less instructive in
assessing the DL model versus the hst1pass one, but include it for
the sake of completeness, as it was also presented in Paper I.
\subsection{Characterizing the Impact of Global Positions
  and Magnitude \label{subsec:impact}}
One of the main modifications made to our DL model, over that presented
in Paper I, is the inclusion of layers making use of the star's rough 
global position and instrumental magnitude.
The expectation is that these would effectively model the known variation 
of the PSF across the chip and any possible dependence on magnitude.
How important are these new parameters in the derived DL model solutions?

As a trained DL model solution can very much be a 'black box' in some respect,
a simple test was performed to measure the sensitivity of the model output
to these specific input parameters.
A single star image was chosen from an exposure in the NGC 104 data set, more
precisely, one star near the center of each chip for a single F555W exposure and
a single F814W exposure.
The star images were selected to be relatively bright, that is, with high signal
to noise.
The DL model solution, appropriate to that chip and filter, was applied to the
star's image, using its actual intensity raster cutout and correct global position 
and instrumental magnitude.
Additionally, artificial repeats of the same image raster cutout were made placing
it at a grid of positions across the chip, while keeping the input magnitude
correct.
Finally, repeats were also made at the correct, near-center position, but with a
range of input magnitudes, spanning the effective range of magnitudes used to
train the model, i.e., the range over which the model solution should be legitimate.

The changes in the DL-calculated output centers, as a function of global position 
and magnitude, will measure the dependence of the centers on those input parameters.
The calculated center at the actual position and magnitude is used as fiducial and
differences are taken with the centers derived while varying the global position
and magnitude, taking into account the additional integer-value offsets for the
grid of global positions.

\begin{figure}[H] 
\includegraphics[scale=0.52,angle=-90]{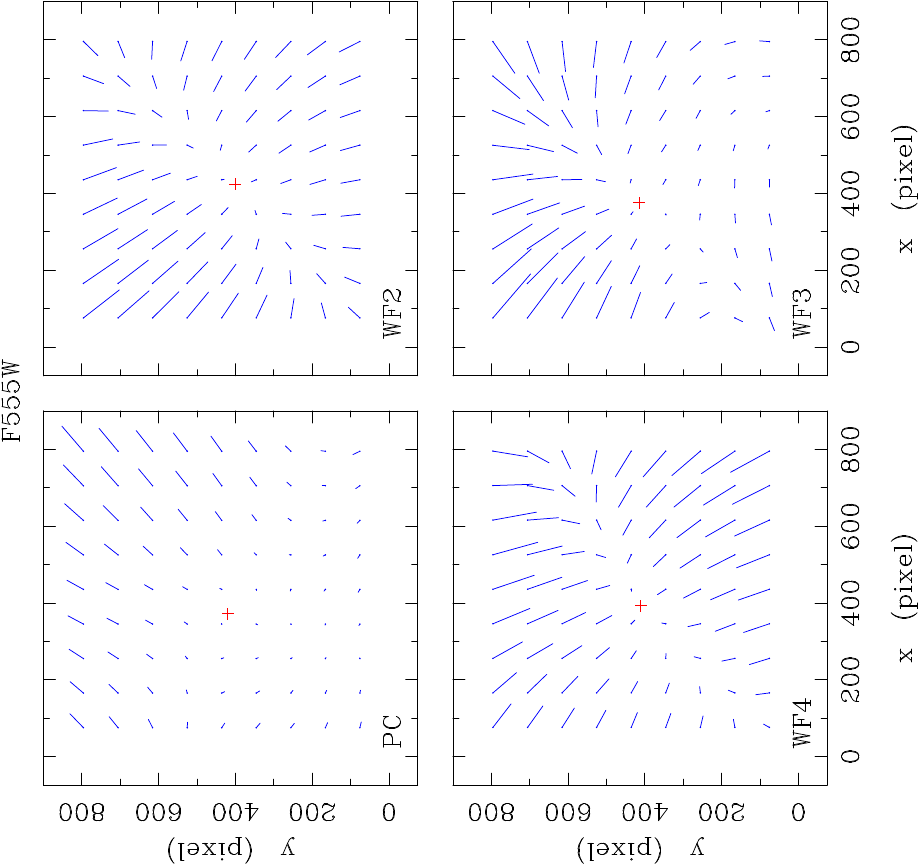}
\caption{Sensitivity of the trained DL models to location on the chip, for
an F555W exposure.
At each 2-d grid point, the vector shows the amount of change in
model-calculated $x$ and $y$ star-image position had this centrally
located star's image actually been at the indicated location on the chip.
The vectors are magnified by a factor of 1000.
The length of the longest vector shown is roughly 160 mpix.
The plus symbol indicates the actual location of the star image on the chip.
The four WFPC2 chips are as labeled.
}
\label{fig:f5-vec}
\end{figure}
\begin{figure} [H]
\includegraphics[scale=0.52,angle=-90]{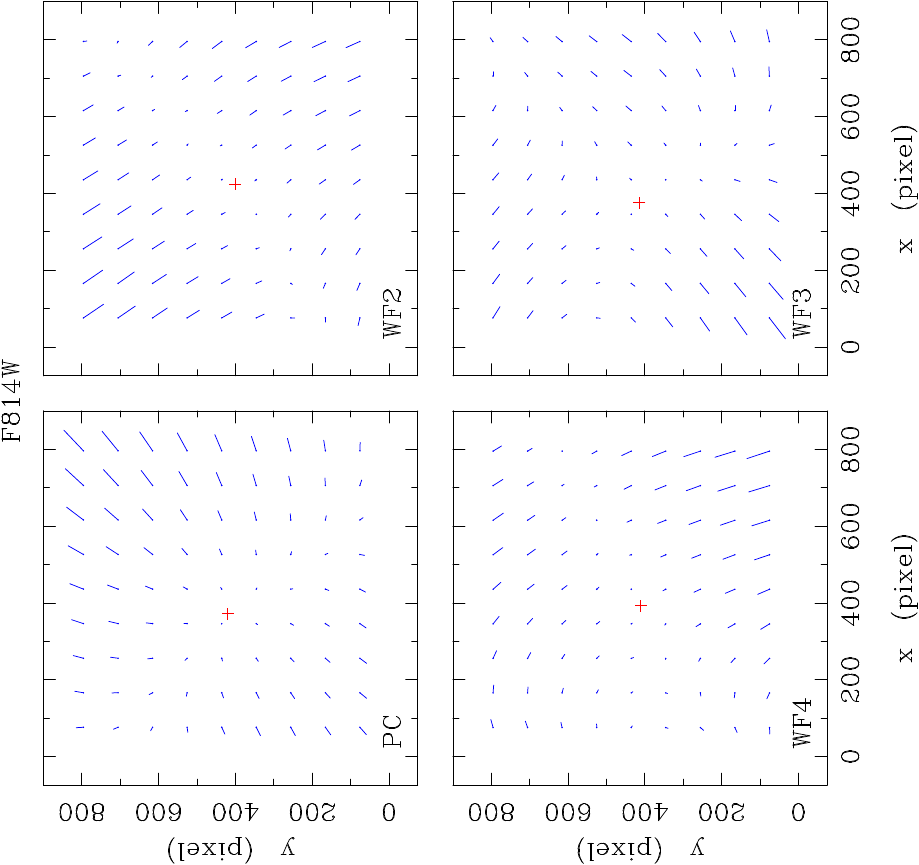}
\caption{The same as Fig \ref{fig:f5-vec} but for a filter F814W exposure.}
\label{fig:f8-vec}
\end{figure}

Figures \ref{fig:f5-vec} and \ref{fig:f8-vec} show the sensitivity of the
calculated positions to the location of the image on the chip.
The vectors indicate the change in calculated image center relative to what
would have been determined had the image been at its true location
near the center of the chip.
The length of the largest vectors in these figures is roughly 160 mpix,
although near the edges of the chip, the changes are more typically of the
order of $\sim 50$ mpix.
This demonstrates a rather large dependence on global position, considering the
$\sim 10$ mpix precision of the centers.
In other words, the PSF does vary significantly across the chip and yet the model
solution is able to make these large adjustments with precision.
Note that the variation, especially for the WF chips, is much greater for the
F555W filter than in F814W.

\begin{figure}[H]
  \plottwo{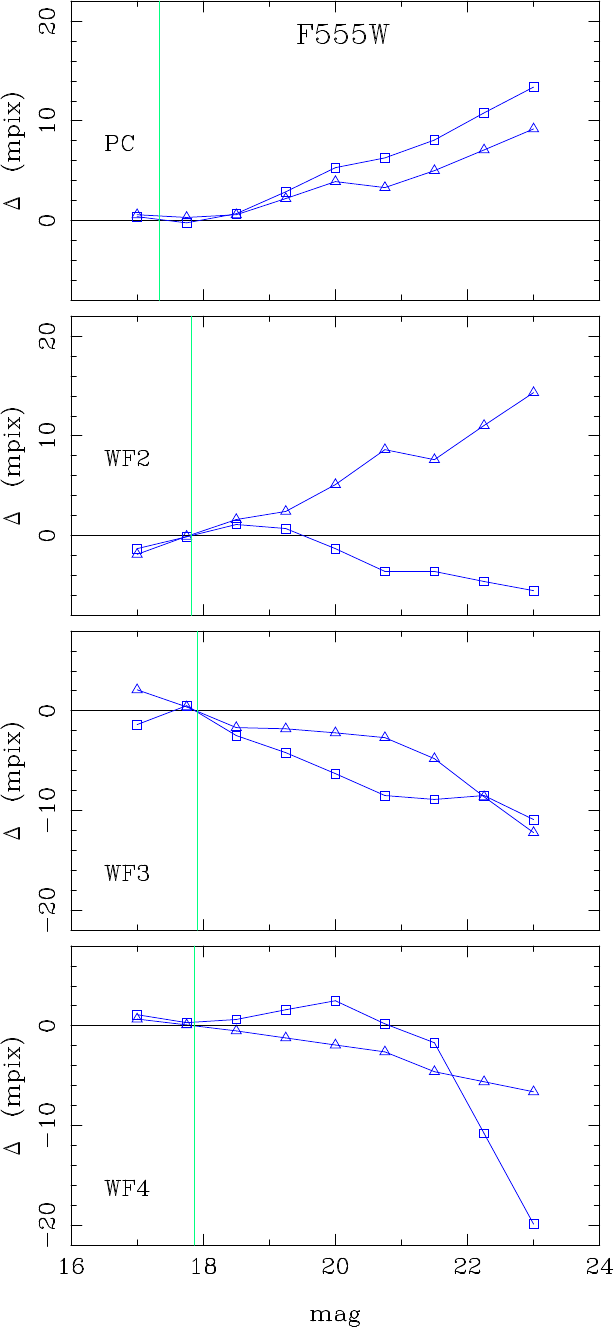}{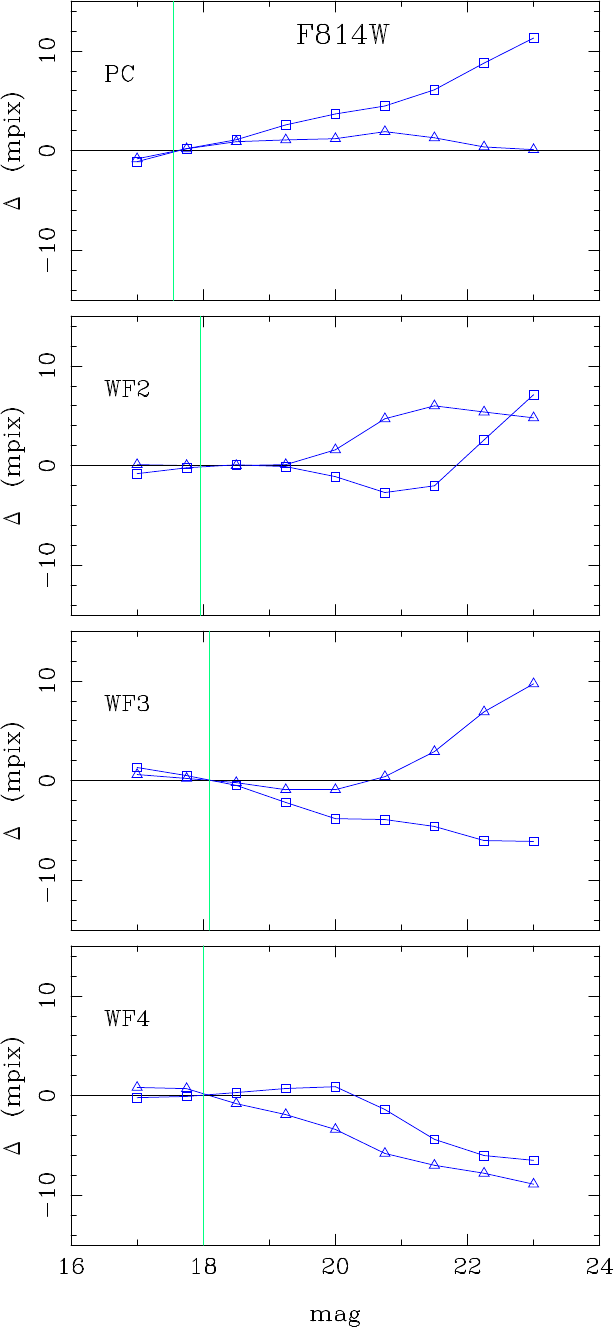}
  \caption{Sensitivity of the trained DL model to input magnitude.
Differences in model-calculated centers are shown as a function of input magnitude.
In each panel, square symbols show changes along the x axis, while triangles show 
the y-axis changes.
The vertical line indicates the actual instrumental magnitude of the star image.
Filter and WFPC2 chip are as labeled.}
\label{fig:f-mag}
\end{figure}

In Figure \ref{fig:f-mag}, the dependence on magnitude is shown.
Within each panel, the two curves illustrate the changes along the $x$ and $y$
axes, while the vertical line indicates the actual instrumental magnitude
of this star's image.
Overall, the dependence on magnitude is of smaller amplitude and is less 
well-behaved, compared to the variation with global position.
Still, the effect of varying the input magnitude over a 6-magnitude
range results in a change in the calculated position of from 10 to 20 mpix.
The sensitivity of the trained DL model to global position is quite large, 
while the effect due to magnitude is smaller but not negligible.

\section{Summary  \label{sec:summary}}
We present a DL methodology that provides improved astrometric
centering for the full field of view of the WFPC2, largely by overcoming the
pixel-phase bias present in these undersampled images.
This bias can be as large as 40 mpix when classic
centering algorithms are used, presumably due to a
mismatch between an algorithm's fitting PSF and the actual PSF.
The procedure we develop relies on a stellar-rich set of repeated
exposures that have small offsets, well-sampled in fractional
pixel phase, for the purpose of training a supervised DL model.

Our new results indicate errors of the order of 8 to 10 mpix in the
centers of well-measured stars.
We also found that the PSF variations across each chip correspond to
corrections of the order of $\sim 100$ mpix, while magnitude effects
are at a level of $\sim 10$ mpix.

While this procedure was developed specifically for
undersampled WFPC2 images,
preliminary testing has shown us that ACS/WFC
exposures in narrow filters
(e.g., F502N) also are affected by pixel-phase
bias when using classic centering
techniques.  Thus, it may very well be possible
to improve ACS/WFC star centering
with a similar DL approach.  Our goal is to
explore this possibility in the future.

\acknowledgments

This work was supported in part by the NASA Connecticut Space
Grant Consortium faculty research
grant 80NSSC20M0129, and by program HST-AR-15632
provided by NASA through a grant from
the Space Telescope Science Institute,
which is operated by the Association of
Universities for Research in Astronomy, Inc.

RBG and ACR are funded by the research Project
``ADELA: Aplicaciones de
Deep Learning para Astrof{\'i}sica'',
with reference PP-2022-13, awarded by the Call
``Proyectos Propios de Investigacion UNIR 2022''.
RBG acknowledges the Call for Grants for Research
Stays Abroad 2022/2023 from UNIR, and the Call
``Estancias de movilidad en el extranjero Jos{\'e}
Castillejo para j{\'o}venes doctores'',
with ref. CAS22/00094, from the Spanish Ministry for Universities.

All the {\it HST} data sets used in this paper can be found in MAST.
The data set number is that from the last column of Table \ref{tab:data-sets} as follows: \\
Set 1: \dataset[http://dx.doi.org/10.17909/n9b8-3721]{http://dx.doi.org/10.17909/n9b8-3721} \\
Set 2: \dataset[http://dx.doi.org/10.17909/n07m-jb63]{http://dx.doi.org/10.17909/n07m-jb63} \\ 
Set 3:\dataset[http://dx.doi.org/10.17909/q64r-dn92]{http://dx.doi.org/10.17909/q64r-dn92} \\
Set 4:\dataset[http://dx.doi.org/10.17909/0d26-v967]{http://dx.doi.org/10.17909/0d26-v967} \\
Set 5:\dataset[http://dx.doi.org/10.17909/mk0b-mx64]{http://dx.doi.org/10.17909/mk0b-mx64} \\
Set 6:\dataset[http://dx.doi.org/10.17909/rhae-wf34]{http://dx.doi.org/10.17909/rhae-wf34} \\
Set 7:\dataset[http://dx.doi.org/10.17909/mzj9-0y54]{http://dx.doi.org/10.17909/mzj9-0y54} \\
Set 8:\dataset[http://dx.doi.org/10.17909/h7dk-6j75]{http://dx.doi.org/10.17909/h7dk-6j75} \\
Set 9:\dataset[http://dx.doi.org/10.17909/9jm2-cf66]{http://dx.doi.org/10.17909/9jm2-cf66} \\

\vspace{5mm}
\facilities{{\it HST} (WFPC2), MAST}


\bibliography{ms}{}

\begin{thebibliography}{}
\expandafter\ifx\csname natexlab\endcsname\relax\def\natexlab#1{#1}\fi

\bibitem[{{Anderson}(2022)}]{and2022}
{Anderson}, J. 2022, {One-Pass HST Photometry with hst1pass}, Instrument
  Science Report ACS 2022-02, ,

\bibitem[{{Anderson} \& {King}(1999)}]{and1999}
{Anderson}, J., \& {King}, I.~R. 1999, \pasp, 111, 1095

\bibitem[{{Anderson} \& {King}(2000)}]{and2000}
---. 2000, \pasp, 112, 1360

\bibitem[{{Anderson} \& {King}(2003)}]{and2003}
---. 2003, \pasp, 115, 113

\bibitem[{{Baena-Gall{\'e}} {et~al.}(2023){Baena-Gall{\'e}}, {Girard},
  {Casetti-Dinescu}, \& {Martone}}]{baena2023}
{Baena-Gall{\'e}}, R., {Girard}, T.~M., {Casetti-Dinescu}, D.~I., \& {Martone},
  M. 2023, in Highlights on Spanish Astrophysics XI, 312

\bibitem[{{Casetti-Dinescu} {et~al.}(2023){Casetti-Dinescu}, {Girard},
  {Baena-Gall{\'e}}, {Martone}, \& {Schwendemann}}]{casetti2023}
{Casetti-Dinescu}, D.~I., {Girard}, T.~M., {Baena-Gall{\'e}}, R., {Martone},
  M., \& {Schwendemann}, K. 2023, \pasp, 135, 054501

\bibitem[{{Chollet} \& {others}(2018)}]{chollet2018}
{Chollet}, F., \& {others}. 2018, {Keras: The Python Deep Learning library},
  Astrophysics Source Code Library, record ascl:1806.022, , , ascl:1806.022

\bibitem[{{Li} {et~al.}(2016){Li}, {Jamieson}, {DeSalvo}, {Rostamizadeh}, \&
  {Talwalkar}}]{Li2018}
{Li}, L., {Jamieson}, K., {DeSalvo}, G., {Rostamizadeh}, A., \& {Talwalkar}, A.
  2016, arXiv e-prints, arXiv:1603.06560

\bibitem[{{Simonyan} \& {Zisserman}(2014)}]{simonyan14}
{Simonyan}, K., \& {Zisserman}, A. 2014, arXiv e-prints, arXiv:1409.1556

\end{thebibliography}

\end{document}